# Measuring the disruptiveness of conceptual papers in the field of marketing

**Jennifer JooYeon Lee[1†*], Hyunuk Kim[2*]**

[1] Department of Administrative Sciences, Metropolitan College,
Boston University, Boston, MA 02215, USA, https://orcid.org/0000-0002-2853-1183

[2] Department of Management and Entrepreneurship,
Martha and Spencer Love School of Business, Elon University, Elon, NC 27244, USA,
https://orcid.org/0000-0002-4863-4729

[†]Corresponding author: leejen@bu.edu

## Abstract

Marketing scholars have underscored the importance of conceptual articles in providing theoretical foundations and new perspectives to the field. This paper supports the argument by employing two network-based measures – the number of citations and the disruption score – and comparing them for conceptual and empirical research. With the aid of a large language model, we classify conceptual and empirical articles published in a substantial set of marketing journals. The findings reveal that conceptual research is not only more frequently cited but also has a greater disruptive impact on the field of marketing than empirical research. Our paper contributes to the understanding of how marketing articles advance knowledge through developmental approaches.

## Summary statement of contribution

We contribute to the marketing literature by expanding on previous discussions of the value of conceptual papers, specifically by collecting and classifying a large set of bibliometric data with the aid of a Large Language Model, along with a rigorous four-stage validation process. Moreover, our practical implications for marketing researchers include the application of an emerging metric (i.e., disruption score) to assess the conceptual papers' potential to disrupt the field, emphasizing the significance of publishing groundbreaking research.

## Introduction

Academic knowledge is developed, disseminated, and utilized to advance the field of marketing (AMA Task Force 1988). Scholarly publications unfold the advancement by motivating the development of new thought, facilitating the dissemination process, and impacting constituencies, including academia and practitioners. Each of the three stages may influence the others; certain methodologies used to develop a research article may result in its efficient dissemination, in turn impacting how much the findings of a study can be generalized (i.e., external validity). Research that matters will only be found when the understanding of the three stages precedes it.

Marketing researchers have sought to identify optimal strategies for producing scholarly work that develops and disseminates knowledge effectively. Some have initiated discussions on different types of methodologies to develop an article (e.g., classification of descriptive, data-driven, empirical, and conceptual articles by MacInnis 2004). Others used bibliometrics to examine how knowledge is emanated to and utilized by the readers. For instance, the impact factor based on citation counts has been conventionally used to measure how well knowledge is disseminated in the field (Bettencourt and Houston 2001; MacInnis 2011; Stremersch, Verniers, and Verhoef 2007; Tellis, Chandy, and Ackerman 1999).

However, the assessment of scholarly impact solely through citation counts is limited in its scope, as it primarily focuses on the three aspects: paper, journal and authorship (Tahamtan, Safipour Afshar, and Ahamdzadeh 2016). This metric often falls short of capturing a paper's potential to open new research directions. In response to this constraint, an alternative methodology has been proposed and used to measure the extent to which a paper consolidates or disrupts a field by using the structure of citation networks (Park, Leahey, and Funk 2023; Wu, Wang, and Evans 2019).

In this study, we aim to understand the types of marketing articles that effectively stimulate knowledge development and dissemination by 1) using a proprietary language model to classify marketing papers based on their knowledge development methods: conceptual and empirical and 2) analyzing the patterns of knowledge dissemination of conceptual and empirical papers with respect to how they impact and disrupt the field. This paper is organized as follows. Firstly, we integrate the literature on the methodologies of knowledge development to justify our selection of conceptual and empirical paper types for the hypotheses. Next, we discuss the conventional and emerging metrics for evaluating the impact of scholarly articles to rationalize the hypotheses on citation counts and disruption scores. Thirdly, we conduct large-scale empirical analyses, along with a rigorous validation process to support our usage of the language model. Finally, we provide the study results on the trend and value of marketing articles.

## Literature Review and Hypotheses Development

### *Methodologies of Knowledge Development: Conceptual versus Empirical Research*

Marketing articles have employed a variety of methodologies, writing styles, formats, and approaches to develop knowledge. Each article varies in its level of emphasis on theories, empirical analyses, or conceptual frameworks. MacInnis (2004) classified the research endeavors on the basis of two attributes – conceptual and empirical content – yielding four distinctive categories of marketing articles: descriptive articles, data-driven articles, empirical articles, and conceptual articles. Similarly, Stremersch, Verniers, and Verhoef (2007) categorized marketing articles focusing on their adoption of conceptual, empirical, methodological, and analytical approaches.

Among various types of research, the marketing discipline has witnessed precipitous growth in empirical methodologies (Rust 2006). Marketing has continued to be more computationally focused, and empirical methodologies have outgrown conceptual approaches (MacInnis 2011; Rust 2006; Yadav 2010). Some researchers expressed concerns over this trend. MacInnis (2004) suspected that "our field may be developing an empirical bias, with authors, reviewers, and/or editors believing that conceptual papers are less scientific or rigorous than empirical papers." Yadav (2010) also criticized that due to the field's focus on data and sophisticated analytical tools, there has been fragmentations within the discipline, leading to scholars' diminishing ability and interest to undertake comprehensive conceptual inquiries. Lehmann, McAlister, and Staelin (2011) pointed out the marketing field's growing emphasis on analytical rigor. They highlighted the potentially negative consequences of prioritizing analytical rigor, which may be at the expense of relevance, communicability, simplicity of scholarly works, managerial relevance, generalizability, or external validity of a study.

Alongside the critiques of the field's empirical emphasis, the significance of conceptual articles has been a long-discussed topic (AMA Task Force 1988; Grether 1976; MacInnis 2011;

Yadav 2010). Conceptual articles are defined as papers primarily focusing on theoretical development without presenting data and/or analysis for purposes of theory testing (Yadav 2010). Research has shared diverse thoughts on the importance of the field's devotion to theoretical contributions and how academic publications can contribute to conceptual advances (Hambrick 2007). MacInnis (2004) discussed the significant roles of conceptual articles in offering novel prospects and helping us move from a microscopic perspective to a macroscopic one. Other scholars also stressed the importance of marketing discipline to create its own theories of marketing phenomena (Stewart and Zinkhan 2006) and to suggest systematic template-based approaches to developing a conceptual paper (Jaakkola 2020).

Despite the marketing group's consensus on the importance of theories, past studies identified an overall decline in the frequency of marketing conceptual publications (MacInnis 2004; Yadav 2010), possibly because of unfavorable review processes for theory-only articles (Stewart and Zinkhan 2006). In this study, we address this paradox by systematically comparing the scholarly impact of conceptual and empirical articles using two distinct measures: citation counts and disruption score.

***Measures for Knowledge Dissemination***

The ongoing scrutiny of knowledge development methodologies calls for measures to systematically assess publications' potential to be disseminated in the field. Based on Yadav (2010)'s criteria, we select citation counts and disruption scores as the main measures to quantify the impact of conceptual articles in marketing.

Yadav (2010) suggested the five criteria as a guideline to evaluate conceptual articles: exposition, theory building, potential impact, innovativeness, and validity. The author points out that while other elements can be applied in a more straightforward manner, evaluating an article's validity, potential impact and innovativeness presents substantial challenges. The judgments regarding the potential impact and the innovativeness focus on "the promise, the future potential, the problem-solving capacity, or what we might call the 'opportunity profile' of a claim" (Nickles 2006). The future orientation of these elements makes it challenging to assess conceptual articles, yet it is important for evaluating their forward-looking impacts.

Consequently, we focus on the potential impact and innovativeness elements in an effort to understand conceptual papers' potential to open new research directions. Firstly, Yadav (2010)'s potential impact factor represents the scope and significance of issues in the substantive, conceptual, and/or methodological domains that may require reassessment in light of the conceptual arguments. Secondly, the innovativeness element addresses the heuristic power of a research article, which is the potential fruitfulness of the conceptual arguments presented, judged on the basis of forward-looking outcomes. In the forthcoming sections, we first propose employing the conventional measure of citation counts to quantify the potential impact of a conceptual article. Subsequently, we adopt another measure known as the disruption score to capture the innovativeness aspect.

***A Conventional Metric: Citation Counts to Measure the Potential Impact***

Citation counts are widely used for assessing the potential impact of articles (Stremersch, Verniers, and Verhoef 2007). They are used to measure the influence of the work on the scientific community, as high-quality works will trigger more following studies than low-quality works (Bornmann and Daniel 2008). It quantifies the scientific impact of knowledge. Jaakkola and Vargo (2021) extended the discussion by classifying the typical proxies for scientific impact into citation counts, citation patterns within the discipline, citation flows across disciplines, and cross-disciplinary borrowing of theories. Citation counts continue to play a crucial role in important decisions related to hiring, promotion, reputation, funding, and institutional rankings; writing "impactful" articles is a prevailing mantra in academia (Jaakkola and Vargo 2021).

Although previous research has recognized the value of conceptual papers in advancing the marketing field (Lehmann et al. 2011; MacInnis 2011; Stremersch et al. 2007), their actual impact has received limited empirical investigation. A few studies have sought to examine this issue, but their findings remain inconclusive. For example, Yadav (2010), in their comprehensive analysis of thirty years' worth of marketing publications, found that conceptual articles, relative to their numbers, receive a disproportionately high number of citations. This result aligns with Lindsey (1989)'s early criticism of citation counts, arguing that such metrics may underestimate the contributions of applied scientists. In contrast, Stremersch, Verniers, and Verhoef (2007) found no significant relationship between conceptual articles and citation counts. To reconcile these conflicting findings and advance the discussion, our study adopts citation counts as a conventional measure to replicate the previous hypotheses.

**Hypothesis 1**: Conceptual articles have higher citation counts than non-conceptual articles in marketing journals.

Although citation counts have the advantages of being objective and giving credit to broadly relevant papers, they are not without biases. Lehmann, McAlister, and Staelin (2011) criticized that citation counts are weighted toward topics that were relevant in the past. Citation counts also tend to favor review or new method types of papers over others and only capture direct references. Additionally, they may be subject to "game playing," where authors strategically cite the influential individuals they think, or hope, will review the paper. According to Koltun and Hafner (2021), the correlation between the h-index and recognition awards granted by the scientific community has substantially declined, which signals that citation counts no longer effectively measure the scientific reputation. Shakeel et al. (2022) also criticize the meaning, value, and comparability of citation counts, suggesting alternative metrics.

***An Emerging Metric: Disruption Score to Measure the Innovativeness Element***

The marketing field has increasingly recognized the importance of research that challenges established norms and institutions (Jaakkola and Vargo 2021). Such work often contributes to novel and disruptive advancements—dimensions of impact that citation counts alone fail to fully capture (Wu, Wang, and Evans 2019; Wang, Zhou, and Zeng 2023; Leibel and Bornmann 2024). To address this limitation, our research quantifies the disruptiveness of marketing articles by analyzing their citation patterns. Specifically, we adopt a variant of the *CDt* index proposed by Funk and Owen-Smith (2017), who define disruptiveness as the extent to which an object consolidates or destabilizes the streams of advancements. We refer to their index to reduce the disproportionate influence of highly cited papers in the calculation, following the recent discussions in scientometrics (e.g., Bornmann et al. 2020).

The disruptiveness quantified by the *CDt* index has been widely used to examine various factors of knowledge development. Sheng et al. (2023) found complex relationships between the disruptiveness and features of the prior knowledge of a paper. They found that the number of references of a focal paper and the average citations of these references are positively associated with the disruptiveness of the focal paper. On the other hand, the recency of references was found to be negatively associated with the disruptiveness. Kedrick et al. (2024) showed that academic papers tend to be more disruptive if their field has a larger core of concepts and this core has a higher churn. This tendency was found in the social sciences and physical sciences domains.

The community of marketing scholars initiated the discourse concerning novel and disruptive research, placing greater value on conceptual contributions (Clark et al. 2014; MacInnis 2011; Yadav 2010), research addressing high-relevance topics (Kohli and Haenlein 2021; Reibstein, Day, and Wind 2009), and interdisciplinary work (Moorman et al. 2019; Yadav 2018; Chen et al. 2024). In response to ongoing discussions about the role and novelty of conceptual works, the present work empirically measures the extent to which conceptual research in marketing contributes to disruption and innovation in the field.

Hypothesis 2 builds on the foundation laid by Hypothesis 1 by exploring the relationship between article type and disruptiveness within the marketing domain. While previous scientometric studies have examined disruption and citation dynamics, this study uniquely applies those insights to contrast the disruptive potential of conceptual versus non-conceptual articles in marketing. Notably, Wang et al. (2023) identified a weak correlation between disruption and citation counts, underscoring the need to examine disruptiveness as a distinct dimension of scholarly impact.

**Hypothesis 2**: Conceptual articles have higher disruption scores than non-conceptual articles in marketing journals.

A valid classification of conceptual versus non-conceptual articles should precede the testing of the hypotheses. The following sections will discuss our classification process using GPT, as well as the validation process to support the use of this tool.

## Methods

### *Using GPT to Identify Conceptual Articles*

GPT has great potentials for text annotation tasks as its performance is comparable to human experts and crowd workers, as evidenced in previous studies on identifying hate speeches (Huang, Kwak, and An 2023), classifying political affiliations from Twitter posts (Törnberg, 2023), and detecting the semantic characteristics of Twitter posts (Gilardi, Alizadeh, and Kubli 2023). We borrowed the definitions of conceptual and empirical articles from the editorial guidelines of the Journal of Marketing[1] and ask GPT-4o mini to classify an article into the conceptual or the empirical category based on its title and abstract by the definitions. The temperature of GPT-4o mini is set to 0 to get the most deterministic outcomes. The GPT query was input as below. “%s” in the query is the place for texts.

[1] https://www.ama.org/editorial-guidelines-journal-of-marketing/

*There are two types of articles published in the Journal of Marketing as below.*
*1. Conceptual articles: These types of articles make their contributions through theoretical arguments that introduce new topics, new constructs, new relationships, new theories, and even new paradigms for the field.*
*2. Empirical articles: Empirical articles use organized observations about marketing-relevant data of any type to offer important insights to the marketing discipline.*
*Based on the definitions above, classify an academic article with title “%s” and abstract “%s” into either the conceptual category or the empirical category.*

To ensure whether GPT can be used to identify the article types, we compared GPT responses with the well-known lists of conceptual and empirical articles from the marketing journals. The overall accuracy of capturing conceptual articles is found to be over 90% across all four validation methods below. Although GPT does not attain the cognitive level of marketing scholars, our validations underscore its utility as an efficient tool for this annotation task.

*Validation 1: Journal of the Academy of Marketing Science (JAMS) publications*

JAMS provides the labels for their articles relevant to this research: “Conceptual/Theoretical Paper” and “Original Empirical Research”. From the five issues of JAMS from 2018 to 2022, we collected the titles and abstracts of 28 conceptual/theoretical papers and 214 original empirical research papers. GPT-4o mini correctly classified 25 conceptual/theoretical papers (89.3%) and 207 empirical research papers (96.7%).

*Validation 2: AMS Review publications*

AMS Review is a journal specialized in conceptual contributions to the field of marketing. For 115 publications with the label of “Theory/Conceptual” in all issues, GPT-4o mini correctly classified 105 papers (91.3%) as conceptual articles.

*Validation 3: GPT classifications for the exemplar conceptual articles from Yadav (2010)*

Out of 23 exemplar conceptual articles listed in the paper titled “The Decline of Conceptual Articles and Implications for Knowledge Development” by Yadav (2010), GPT-4o mini correctly classified 21 articles as conceptual articles (91.3%).

*Validation 4: Human annotators*

To further validate our GPT-driven approach, we had five PhD scholars independently classify 30 randomly sampled publications from our regression analysis set into conceptual or empirical categories (See the subsection named 'Ordinary Least Squares (OLS) Regressions'), guided by the Journal of Marketing's established definitions. GPT's classifications showed strong alignment with the scholars' consensus, defined as agreement among at least three of the five experts, matching in 29 out of 30 cases. This demonstrates a high 96.67% classification accuracy rate for GPT and successfully confirms the alignment between expert opinion and the large language model's output.

***Data***

We collected bibliographic information from academic articles using the Dimensions database, which is a large scholarly database covering diverse disciplines (Herzog, Hook, and Konkiel 2020). To limit our analyses to active and data-accessible marketing journals, we selected 98 marketing journals 1) ranked in the first or second quartile of Scimago journal ranking (Category ID: 1406) and 2) indexed in Web of Science 2024. 86,267 articles published between 1991 and 2020 in these journals were retrieved from the database.

For each focal article from this set, we then obtained its references and all papers citing either the focal article itself or any of its references within a five-year window of the focal article's publication. Each citation establishes a directional link; for example, if a paper $i$ is cited by $p$, a link $i \rightarrow p$ is created.

***Disruption Measure***

To quantify the extent to which academic articles disrupt the field of marketing and lead subsequent research, we calculated their disruption scores by adopting a network-based measure suggested as the '*CDt* Index' (referred to as $D$ in this paper) by Funk and Owen-Smith (2017) for patents, which was later used by Wu, Wang, and Evans (2019) for academic papers. As mentioned in the previous section, we only considered citations within a five-year window. $D_i$, the disruption score of an article $i$, is defined as $(N_F - N_B)/(N_F + N_B + N_R)$ where $N_F$ is the number of articles citing only $i$, $N_B$ is the number of articles citing both $i$ and $i$'s references, and $N_R$ is the number of articles citing only $i$'s references. The higher an article's $D$ is, the more the article disrupts the existing line of research. Figure 1 shows two examples. In the left network of Figure 1, $N_F$ is 3 ($p_1, p_2, p_3$), $N_R$ is 1 ($p_4$), and $N_B$ is 0, leading $D_i$ to be 3/4. In the right network of Figure 1, $N_F$ is 1 ($p_2$), $N_R$ is 1 ($p_4$), and $N_B$ is 2 ($p_1, p_3$), leading $D_i$ to be -1/4.

Figure 1: Examples of the disruption scores of an article $i$. The node labeled $r$ indicates an article cited by $i$, and the node labeled $p$ indicates an article citing $i$ or $i$'s references.

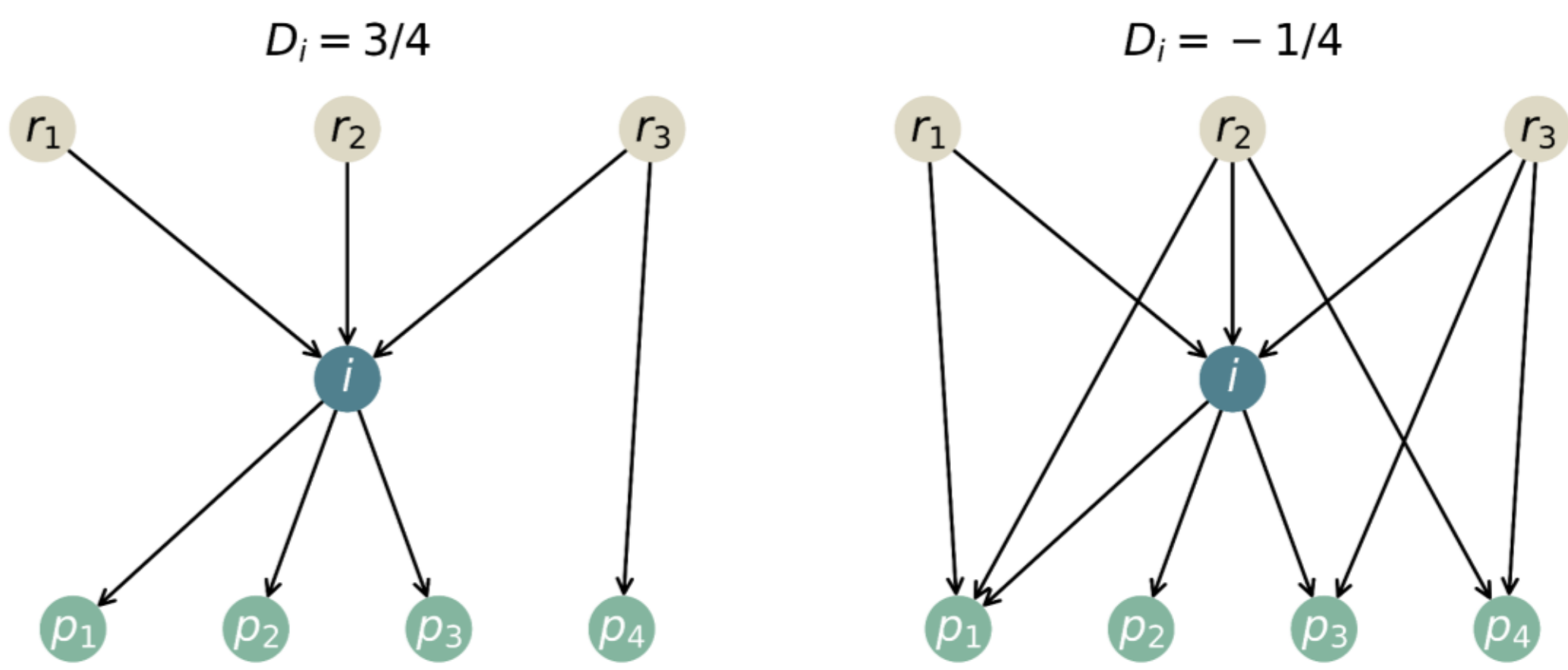

This index well captures the disruptive entities in a network but is not suitable for some disciplines where many research articles tend to rely on some highly cited, seminal articles. The field of marketing is not an exception. For instance, Thaler (1985)'s seminal work on mental accounting, which was known as the most cited article in Marketing Science as of 2002 (Shugan 2002), continued to gain attention in the field and maintained its position as the second most cited article in the year of 2017 when Thaler received the Nobel Prize in Economics (Sudhir 2017). In such areas where emerging articles tend to cite strong, seminal articles, the disruption score of a new article could be low even if it contains disruptive knowledge. To address this issue, we adopted a variant proposed by Bornmann et al. (2020), which applies a threshold $l$ to count articles ($N_B$) citing the focal paper and at least $l$ references of the focal paper. We term this variant ${N_B}^l$, and then $D^l$ is equal to$(N_F - {N_B}^l)/(N_F + {N_B}^l + N_R)$. If $l$ is set to 2 for the network on the right hand side of Figure 1, the disruption score becomes 0 as $N_F$ is 1 ($p_2$), $N_R$ is 1 ($p_4$), and ${N_B}^l$ is 1 ($p_1$) because $p_1$ cited two references of $i$, $r_1$ and $r_2$. Bornmann et al. (2020) recommended $l = 5$, and we report results for $l = 1,3,5$ where $D^1$ is the original CD index. These values were calculated for 86,179 articles.

***Ordinary Least Squares (OLS) Regressions***

To ensure sufficient textual input for GPT's classification, we included 82,992 articles with abstracts exceeding 500 characters (including whitespaces) in our regression analyses. GPT classified these into 13,072 conceptual and 69,920 empirical articles. Our results remained consistent across different abstract length thresholds (see Supplementary Information).

We set the dependent variables as the number of citations and disruption scores calculated from citations within the 5-year window. We explained these using five variables: publication year group (5-year group), the number of authors, a binary indicator of whether an article is conceptual (referred to as 'Conceptual' in regression tables), Scimago Journal Ranking (SJR) quartile (Q1 or Q2 – Q1 is used as a reference group so only Q2 is included in regression tables), and a binary indicator of whether an article acknowledges a grant (referred to as 'Grant' in regression tables). The publication year group is included to check if older articles tend to have higher numbers of marketing citations and if the declining trend of disruptive knowledge over time (Park et al. 2023) is also found in the field of marketing. The number of authors is included to check if large teams are associated with the articles with higher numbers of marketing citations and if small teams are more likely to conduct disruptive research as Wu, Wang, and Evans (2019) found in the science and technology disciplines. The SJR quartile was included to assess whether journal quality affects the citation impact and disruptiveness. Similarly, the Grant variable was added because grant funding is a recognized correlate of research impact (Rigby 2013).

## Results

### *Citation Impact of Conceptual Articles*

Our four OLS regression models (Table 1) confirm that conceptual articles in marketing tend to have received more citations than empirical articles, supporting Hypothesis 1. Specifically, Model 1 shows that academic articles were cited more in recent years. Model 2 further shows that the number of authors is significantly associated with the citation counts, aligning with existing literature (Tahamtan, Safipour Afshar, and Ahamdzadeh 2016). After accounting for these factors, Model 3 demonstrates that conceptual articles exhibit higher citation counts. This positive and significant citation impact of conceptual articles persists in Model 4, even after controlling for SJR quartile and grant acknowledgment.

Table 2: The OLS regressions explaining the number of citations with publication year groups, the number of authors, a binary indicator of whether an article is conceptual, Scimago Journal Rank quantile (Q1 or Q2), and a binary indicator of whether a grant is associated with.

| | Dependent Variable: The Number of Citations | | | | | | | |
|---|---|---|---|---|---|---|---|---|
| | Model 1 | | Model 2 | | Model 3 | | Model 4 | |
| | Coef. (SE) | *p* | Coef. (SE) | *p* | Coef. (SE) | *p* | Coef. (SE) | *p* |
| Intercept | 6.061 | < .001 | 1.208 | .144 | -1.675 | .045 | 2.976 | < .001 |
| 1996-2000 | 1.754 | .076 | 1.571 | .112 | 1.601 | .104 | 2.936 | .003 |
| 2001-2005 | 6.606 | < .001 | 6.052 | < .001 | 6.284 | < .001 | 7.605 | < .001 |
| 2006-2010 | 10.45 | < .001 | 9.541 | < .001 | 9.883 | < .001 | 11.20 | < .001 |
| 2011-2015 | 15.29 | < .001 | 13.83 | < .001 | 14.38 | < .001 | 15.74 | < .001 |
| 2016-2020 | 33.38 | < .001 | 31.29 | < .001 | 31.89 | < .001 | 33.72 | < .001 |
| # of Authors | | | 2.478 | < .001 | 2.837 | < .001 | 2.577 | < .001 |
| Conceptual | | | | | 10.02 | < .001 | 10.42 | < .001 |
| SJR Q2 | | | | | | | -17.14 | < .001 |
| Grant | | | | | | | 1.742 | .032 |
| # of Observations | 82,992 | | 82,992 | | 82,992 | | 82,992 | |
| Adjusted R-squared | .052 | | .056 | | .060 | | .084 | |

***Disruptiveness of Conceptual Articles***

As we hypothesized, conceptual articles impact the marketing academia not just by being frequently cited by the subsequent papers, but by opening new branches of research (Hypothesis 2 supported). We ran OLS regressions for $D^1$, $D^3$, and $D^5$ with the same variables (Table 2). For $D^1$, which is the original CD index while still sensitive to highly cited papers, 'Conceptual' is marginally significant. On the other hand, for $D^3$ and $D^5$, 'Conceptual' becomes significant at the level of 0.001. Similar to the Park et al. (2023)'s finding of the decline in disruptive knowledge over time in the science and technology fields, the coefficients of the publication year groups are found to decrease over time, according to the models for $D^3$ and $D^5$. The number of authors is negatively associated with disruption scores, suggesting large teams tend to develop the field of marketing by consolidating existing research streams, whereas small teams do so by introducing new research directions.

Table 3: The OLS regressions explaining $D^1$, $D^3$, and $D^5$ with publication year groups, the number of authors, a binary indicator of whether an article is conceptual, Scimago Journal Rank quantile (Q1 or Q2), and a binary indicator of whether a grant is associated with.

| | Dependent Variable | | | | | |
|---|---|---|---|---|---|---|
| | $D^1$ | | $D^3$ | | $D^5$ | |
| | Coef. (SE) | *p* | Coef. (SE) | *p* | Coef. (SE) | *p* |
| Intercept | .029 | < .001 | .041 | < .001 | .044 | < .001 |
| 1996-2000 | -.008 | < .001 | -.011 | < .001 | -.012 | < .001 |
| 2001-2005 | -.014 | < .001 | -.020 | < .001 | -.021 | < .001 |
| 2006-2010 | -.024 | < .001 | -.032 | < .001 | -.034 | < .001 |
| 2011-2015 | -.025 | < .001 | -.035 | < .001 | -.037 | < .001 |
| 2016-2020 | -.026 | < .001 | -.036 | < .001 | -.038 | < .001 |
| # of Authors | -.001 | < .001 | -.002 | < .001 | -.002 | < .001 |
| Conceptual | .005 | < .001 | .006 | < .001 | .006 | < .001 |
| SJR Q2 | .001 | .240 | .001 | .015 | .001 | .010 |
| Grant | -.001 | .226 | -.001 | .326 | -.001 | .418 |
| # of Observations | 82,992 | | 82,992 | | 82,992 | |
| Adjusted R-squared | .017 | | .029 | | .032 | |

## Conclusion

### *Discussion*

Our research indicates that marketing conceptual articles exhibit a higher level of disruptiveness compared to empirical ones. Recognizing the potential of conceptual articles to disrupt the marketing field, we suggest the imperative need to develop and disseminate conceptual articles as a means to advance academia.

There are two hurdles to attaining academic disruption, especially considering the surging quantity of scholarly publications: 1) new publications' limited chance against the seminal papers, and 2) conceptual papers' challenges in the presence of empirical overemphasis. Firstly, the overall increasing number of publications can slow down the development and recognition of new ideas, which may be exacerbated as citations are more likely to be made to the seminal papers published in the earlier years (Chu and Evans 2021). Secondly, previous research has presented concerns over the outgrowth of empirical methodologies in marketing (MacInnis 2011; Rust 2006; Yadav 2010). Our data also shows that the empirical articles have been more prevalent than conceptual articles in terms of publication quantity, which may further limit the advancement of conceptual publications. It is worthwhile to note that despite this imbalanced number of publications, we find that conceptual articles are cited more frequently than empirical articles, which demonstrates their capability to impact the field.

The field's continued endeavor to highlight the value of new conceptual articles may help us address such considerations. For instance, we suggest the leading academic societies publish special issues on conceptual ideas or create niche journals or venues specialized in conceptual articles, such as AMS Review, a journal that is positioned to focus exclusively on conceptual contributions. By doing so, we can potentially broaden the scope of the field of marketing and clear some impediments to academic disruptions.

***Theoretical and Practical Implications***

As discussed earlier, academic knowledge is developed, disseminated, and utilized to advance the field (AMA Task Force 1988). The three stages form an interconnected relationship where each stage dynamically influences the others. For instance, how knowledge is developed may impact how efficiently the idea is disseminated in the field, which may further determine if the idea is practically utilized by the recipients.

The present research contributes to theory by advancing the understanding of the earlier parts of this knowledge process: development and dissemination of knowledge. Our findings yield implications for two of the knowledge development methodologies (i.e., conceptual and empirical) and the two measures to assess the potential of knowledge dissemination (i.e., citation impact and disruption score). Specifically, we contribute to the marketing academia by expanding the previous discussions on the value of conceptual papers (e.g., MacInnis 2004; Yadav 2010). We extend the previous findings on the conceptual papers' citation impact by collecting additional data and identifying the articles published in the recent decade. Moreover, our practical implications for marketing researchers include an application of the novel metric to measure the conceptual papers' potential to disrupt the field. Further discussion on the disruption score may contribute to the marketing researchers' recognition of the significance of publishing innovative and groundbreaking research. Additionally, we promote the practical usage of the emerging GPT tool to efficiently classify a large set of data. Our four-stage validation process also provides a robust foundation for future GPT-based practices to build upon.

***Limitations and Future Research***

We discuss a few limitations of the present research. Firstly, our research is limited in providing implications for the last part of the knowledge process, the utilization stage. This stage relates to scholars' motivation to conduct research that matters and research that influences society. It is worth noting that Vargo and Koskela-Huotari (2020) raised a potential issue regarding the conceptual articles' limited ability to provide practical impact. They state that "this dearth of conceptual and theoretical articles is leading to a situation in which marketing is becoming characterized as a theory-importing-only discipline and increasingly marginalized," partially due to "conceptual articles not being adequately practical." Accordingly, we suggest further research on conceptual marketing papers' potential to produce knowledge to solve real-life, practical problems, or to make informed decisions. One possible direction is to investigate the academic citation network of business-related industry patents.

Secondly, the methodology employed in the present study comes with inherent limitations. To start, identifying conceptual articles is not a trivial task. Human annotators may read the full texts and determine whether an article is developed through conceptual approaches. While their decisions may be credible, it demands a significant investment of time and effort. On the contrary, employing large language models expedites the process compared to the manual annotation but may come at the expense of accuracy. To enhance the efficacy of the large language models, it is recommended for future research to undertake further comprehensive validations, supplementing the four-stage validations conducted in the present study.

Specifically, we suggest further research to fine-tune the large language models by using the bibliographic information of known conceptual papers as the training data. The latter approach has been successful for classification tasks in the disciplines of law (Liga and Robaldo 2023) and biomedical sciences (Bousselham and Mourhir 2024). These methodological limitations, however, are expected to ease over time; As the capabilities of the large language models continue to evolve (e.g., GPT-4.1), the divergence between human judgment and AI-driven decisions is expected to narrow. This evolution holds the promise of refining our ability to accurately identify conceptual articles.

Lastly, this study focuses specifically on the field of marketing, where the distinction between conceptual and empirical research has been a particularly prominent and evolving topic. We selected the discipline based on its documented debate between the two types of research (e.g., Yadav 2010, Stremersch et al. 2007), as well as the field's evolution from theory-driven foundations to data-intensive subfields, such as data-driven marketing and marketing analytics. Nonetheless, we acknowledge the potential for future research to extend our approach and findings to other business disciplines, including management, finance, and accounting.

**Acknowledgments**

This paper was written using data obtained on April 29, 2025, from Digital Science's Dimensions platform, available at https://app.dimensions.ai.